\documentstyle[manuscript,aps,epsf,eqsecnum]{revtex}

\bibstyle{unsrt}
\begin{document}

\title{Numerical Simulations of ${\cal PT}$-Symmetric Quantum Field Theories}

\author{Claude Bernard\thanks{electronic mail: cb@lump.wustl.edu}
and Van M. Savage\thanks{electronic mail: vmsavage@hbar.wustl.edu}
}
\address{Department of Physics, Washington University, St. Louis, Missouri
 63130, USA}

\date{\today}
\maketitle

\begin{abstract}
Many non-Hermitian but ${\cal PT}$-symmetric theories are known to have a real 
positive spectrum. Since the action is complex for there theories, 
Monte Carlo methods do not apply.  In this paper 
the first field-theoretic method for numerical simulations of ${\cal PT}$-symmetric 
Hamiltonians is presented. The method is the complex Langevin equation, which 
has been used previously to study complex Hamiltonians in statistical physics and in Minkowski space.
We compute the equal-time one-point and two-point Green's functions 
in zero and one dimension, where comparisons to known results can be made.
The method should also be applicable in four-dimensional space-time. 
Our approach may also give insight into how to formulate a 
probabilistic interpretation of ${\cal PT}$-symmetric theories.
\end{abstract}


\section{Introduction}
\label{s0}
Traditionally, only theories with Hermitian Hamiltonians are 
studied in quantum mechanics and quantum field theory.
This is because Hermiticity guarantee real eigenvalues and therefore 
unitary time translation and conservation of  probability. 
It has recently been observed that quantum mechanical theories whose 
Hamiltonians are not Hermitian but are symmetric under a transformation known
as  ${\cal PT}$ symmetry have positive definite spectra
\cite{r0.1,r0.2,r0.3,r0.4,r0.5,r0.6,r0.7,r0.8,r0.9,r0.10,r0.11,r0.12,r0.13,r0.14,rf2,r0.15}. A major
criticism of these theories is that a consistent probabilistic
interpretation has not been formulated. This paper
suggests that there is a real, Fokker-Planck probability underlying these theories
and presents a numerical method for calculating the 
$k$-point Green's functions, $G_k$, of these theories.

A ${\cal PT}$-symmetric Lagrangian that has been studied in the past is
defined by the Euclidean Lagrangian
\begin{equation}
{\cal L}_E={1\over2}(\partial\phi)^2+{1\over2}m^2\phi^2-{g\over N}(i\phi)^N.
\label{eq1.1}
\end{equation}
A recent paper used Schwinger-Dyson techniques to study
this self-interacting scalar quantum field theory \cite{r1}.
Green's functions, $G_k$, calculated by this method agreed extremely 
well with known results. It was argued that these theories possess a positive definite spectrum 
and a nonvanishing value of $G_1=\langle0|\phi|0\rangle$ for all $N>2$. 

Under ${\cal PT}$ symmetry ${\cal P}$ sends $\phi\to -\phi$ and ${\cal T}$
sends $t\to -t$ and $i\to -i$, where $t$ is time. (Note that in one dimension (i.e. quantum mechanics)
$\phi$ represents the position of the particle and ${\cal P}$ corresponds to reflection
in space.) Thus, ${\cal L}_E$ is manifestly ${\cal PT}$ symmetric.
It is believed that the reality and positivity of the spectra are 
a direct consequence of this ${\cal PT}$ symmetry. The positivity of the spectra for all $N$ 
is an extremely surprising result; it is not at all obvious, for example, that the Lagrangian 
${\cal L}_E=(\partial\phi)^2/2-g\phi^4/4$ corresponding to $N=4$ and $m=0$ has a positive spectrum. 
To understand this 
and other results, we must properly define the contours of integration for
path integrals and, in dimension $D>0$, the boundary conditions in the corresponding
Schr\"odinger equation.

Contours of integration and boundary conditions have been extensively studied in zero
and one dimension \cite{r0.1,r1}. In zero dimensions one must choose the contour of integration for the
path integral such that it lies in a region where
$\exp(-S(\phi))$ is damped as $|\phi|\to\pm\infty$ and the path integral converges. 
For massless versions of Eq.~(\ref{eq1.1}), these regions are wedges and are chosen to be analytic continuations of the 
wedges for the harmonic oscillator, which are centered about the negative and positive real axes and 
have angular opening ${\pi\over2}$. For arbitrary $N>2$ the anti-Stokes' lines at the centers
of the left and right wedges lie below the real axis at the angles
\begin{eqnarray}
\theta_{\rm left}&=&-\pi+\left({(N-2)\pi\over 2N}\right),\nonumber\\
\theta_{\rm right}&=&-\left({(N-2)\pi\over 2N}\right).
\label{eq0.3a}
\end{eqnarray}
The opening angle of these wedges is ${\pi\over 2N}$.

Similarly, for one-dimensional versions of Eq.~(\ref{eq1.1}) with $m=0$, the Schr\"odinger 
differential equation is 
\begin{equation}
-\psi''(\phi)-{g(i\phi)^N\over N}\psi(\phi)=E\psi(\phi).
\label{eq0.2}
\end{equation}
In Ref.~\cite{r0.1} it was shown how to continue analytically in the
parameter $N$ away from the harmonic oscillator value $N=2$. This analytic
continuation defines the boundary conditions in the complex-$\phi$ plane. The
regions in the cut complex-$\phi$ plane in which $\psi(\phi)$ vanishes exponentially
as $|\phi|\to\infty$ are once again wedges. These wedges also define
the regions in which $\exp(-S(\phi))$ is exponentially damped and the corresponding path integrals
are convergent in one dimension. Once again, the wedges for $N>2$ were
chosen to be analytic continuations of the wedges for the harmonic oscillator.
For arbitrary $N>2$ the anti-Stokes' lines at the centers
of the left and right wedges lie below the real axis at the angles
\begin{eqnarray}
\theta_{\rm left}&=&-\pi+\left({N-2\over N+2}\right){\pi\over2},\nonumber\\
\theta_{\rm right}&=&-\left({N-2\over N+2}\right){\pi\over 2}.
\label{eq0.3}
\end{eqnarray}
The opening angle of these wedges is ${2\pi\over N+2}$. 

Consequently, expectation values for ${\cal PT}$-symmetric theories can be 
understood as path integrals that have been analytically continued in $N$. 
This analytic continuation deforms the contour from the real axis
for the harmonic oscillator, $N=2$, to contours in the complex-$\phi$ whose endpoints 
lie in wedges where $\exp(-S(\phi))$ is damped as $|\phi|\to\infty$ and the path integral converges.
Defining the complex variable $\phi_C$ to follow any contour whose endpoints lie in the appropriate wedges,
${\cal PT}$-symmetric expectation values of operators $A=A(\phi)$ are given by:
\begin{equation}
{\langle0|A|0\rangle\over\langle0|0\rangle}=
{{\int D\phi_C~A(\phi_C)~e^{-S(\phi_c)}}\over\int D\phi_C~e^{-S(\phi_C)}},
\label{eq0.4}
\end{equation} 
where $S(\phi_C)=\int d^{D}X {\cal L}_E[\phi_C(X)]=\int d^{D}X H[\phi_C(X)]$ is the Euclidean space
action. The most common choice of contour is the one along which $\exp(-S(\phi))$ is
purely damped. This contour is defined as $\phi_C=r\exp(i\theta_L),-\infty\leq r\leq 0$ and 
$\phi_C=r\exp(i\theta_R),0\leq r\leq\infty$, where $\theta_L$ and $\theta_R$ are defined
by Eq.~(\ref{eq0.3a}) or Eq.(\ref{eq0.3}).

As mentioned, another remarkable property of the Lagrangian in Eq.~(\ref{eq1.1}) 
is that for all $N>2$ the expectation value
$G_1=\langle0|\phi|0\rangle$ of the position operator $\phi$ in the ground state is nonzero. 
This surprising result shows that the theory is not parity symmetric even when $N$ is even. 
The violation of parity symmetry is a consequence of the manner in 
which the boundary conditions and path integral contours are defined.  
The boundary conditions require that 
as $|\phi|\to\infty$ the anti-Stokes' lines are in the lower half of the complex-$\phi$ plane 
for $N>2$.  While  Eq.~(\ref{eq1.1}) appears to be invariant under a parity 
transformation, the  anti-Stokes' lines are sent to the upper half of the complex-$\phi$ 
plane; this corresponds to a different set of boundary conditions.  Thus, the theory is 
not parity symmetric  except for the special case of $N=2$ where the anti-Stokes' lines 
are the real axis.  Notice that if we now perform a time reversal transformation, complex 
conjugation sends the anti-Stokes' lines back down into the complex-$\phi$ plane, and both 
Eq.~(\ref{eq1.1}) and the boundary conditions are identical to the original formulation 
of the theory.   That is, these theories are ${\cal PT}$-symmetric, but they are not 
symmetric under ${\cal P}$ or ${\cal T}$ separately. As a result, $G_1$ is pure imaginary 
and $G_2$ is real. 

The results for ${\cal PT}$-symmetric but non-Hermitian theories suggest that 
these completely new theories may describe physical processes. Previous
studies have obtained extensive results in zero and one dimension (i.e. quantum mechanics) but have been
unable to perform calculations in higher dimensions. Here, we use the complex Langevin method to calculate 
the equal-time one-point and two-point Green's functions for massless versions of 
Eq.~(\ref{eq1.1}) in zero and one dimension with $N=3$ and $N=4$. The results
are in good agreements with those computed by numerical integration~\cite{r0.1}
and by variational methods~\cite{r0.14,r1}.
Reference~\cite{r9} studies complex Hamiltonians using the Langevin approach in higher dimensions and 
also obtains accurate results. This suggests that the complex Langevin method is the 
robust numerical method needed for studying ${\cal PT}$-symmetric theories in higher
dimensions. Consequently,
we are currently performing field-theoretic, numerical simulations for 
${\cal PT}$-symmetric Hamiltonians in two space-time dimensions. The current paper also
reveals an underlying real Fokker-Planck probability for these theories. We believe our
results represent a significant step towards a physical understanding of ${\cal PT}$-symmetric 
theories.

This paper is organized as follows: In Sec.~\ref{s1} we explain why Monte Carlo
techniques cannot be used for these theories and review the Langevin equation
as a numerical procedure for quantum field theories. In Sec.~\ref{s2} we review the complex Langevin method
and use the methods of supersymmetric quantum mechanics to derive the 
conditions that guarantee convergence for the expectation values.  First and second-order algorithms for 
implementing the Langevin method are presented in Sec.~\ref{s3}, and the results of 
numerical simulations are given and shown to be in excellent agreement with known results. 
Sec.~\ref{s4} contains concluding remarks concerning the implications of this study in 
regard to probability and completeness and proposes higher-dimensional
${\cal PT}$-symmetric theories to which this numerical method could be applied. 

\section{The Langevin Method}
\label{s1}

Since many ${\cal PT}$-symmetric theories possess a real positive spectrum and
a nonvanishing value for $G_1$, it has been speculated that ${\cal PT}$-symmetric theories 
could be used to describe a Higgs Boson. A $-g\phi^4/4$ theory is especially interesting as a theory 
for the Higgs because it has a 
dimensionless coupling constant and is asymptotically free \cite{r1}. However, the Schwinger-Dyson 
equations mentioned in Sec.~\ref{s0} are too difficult to solve in four-dimensional space-time, where 
physical quantities must be calculated. Hence, a reliable numerical method
is needed to compute expectation values in higher dimensions. In this section we argue that 
Monte Carlo methods are ill suited for these theories, while 
the complex Langevin equation provides a robust numerical technique.

Most often, Monte Carlo methods are used for numerical evaluations of expectation
values like those in Eq.~(\ref{eq0.4}) where the contour of integration is along
the real axis and $S(\phi)$ is real. This is achieved 
by choosing paths weighted according to the probability distribution, $\exp(-S(\phi))$. 
For the Hamiltonians considered in this paper, $S(\phi)$ is complex, and neither
a real representation nor a consistent probabilistic interpretation is known. One 
approach to use Monte Carlo is to separate the Hamiltonian into real and imaginary parts 
and consider $A(\phi)\exp(-iIm[S(\phi)])$ the operator and $\exp(-Re[S(\phi)])$ the 
probability distribution. There are two problems with this
approach.  Firstly, the size of $Im[S(\phi)]$ increases with the size of the lattice, making
the numerator and denominator of Eq.~(\ref{eq0.4}) very small. Consequently,
numerical simulations become very difficult. Secondly, and more importantly, in many cases 
$Im[S(\phi)]$ contains more information about which paths are important than $Re[S(\phi)]$. 
Therefore, the algorithm outlined above never samples the important paths and thus fails
to converge to the correct answer. Specific examples of this phenomenon are given in
Refs.~\cite{r3,r4,r5}. Our studies of zero-dimensional versions of Eq.~(\ref{eq1.1}) 
suggest that this is the case for non-Hermitian but ${\cal PT}$-symmetric theories, and that 
Monte Carlo methods are inapplicable.

Another method for numerical calculations in quantum field theory involves
the Langevin equation \cite{r6,r7},
\begin{equation}
{\partial\phi\over\partial\tau}=-{\partial S(\phi)\over\partial\phi} + \eta(\tau),
\label{eq1.3}
\end{equation}
where $\tau$ is an unphysical, Langevin time, $\partial S(\phi)/\partial\phi$ gives
the equations of motion for the Hamiltonian, and $\eta(\tau)$ is a stochastic variable.
The function $\eta(\tau)$ is chosen to be a real, Gaussian random function that
satisfies the conditions 
\begin{equation}
\langle\eta(\tau)\rangle=0,\qquad \langle\eta(\tau)\eta(\tau')\rangle=2\delta(\tau-\tau'),
\label{eq1.4}
\end{equation}
where the averaging is performed with respect to the appropriately normalized 
Gaussian probability distribution. Further, it is well known that when $S(\phi)$
is real, the probability distribution, $P(\phi,\tau)$, associated with Eq.~(\ref{eq1.3}) 
is given by the Fokker-Planck equation \cite{r6,r7}
\begin{equation}
{\partial P(\phi,\tau)\over\partial\tau}={\partial\over\partial\phi}\left({\partial\over\partial\phi}
+{\partial S(\phi)\over\partial\phi}\right)P(\phi,\tau).
\label{eq1.5}
\end{equation}
The space-time dependence of all variables and the $\tau$ dependence of 
$S$ and $\phi$ are left implicit in the above equations. These dependencies are only made 
explicit when relevant to a calculation.

It is easy to evolve  Eq.~(\ref{eq1.3}) numerically in Langevin time, $\tau$, and find 
expectation values. For real variables these expectation values are expressible as 
\begin{equation}
{\langle 0|A|0 \rangle_P\over\langle 0|0 \rangle_P}
={\int{D\phi~A(\phi) P(\phi,\tau)}\over\int{D\phi~P(\phi,\tau)}},
\label{eq1.6}
\end{equation}
and one can show that
\begin{equation}
P(\phi,\tau)\to e^{-S(\phi)},~\tau\to\infty.
\label{eq1.7}
\end{equation}
Hence,
\begin{equation}
{\int{D\phi~A(\phi)P(\phi,\tau)}\over\int{D\phi~P(\phi,\tau)}}\to
{{\int D\phi~A(\phi)e^{-S(\phi)}}\over\int D\phi~e^{-S(\phi)}},~\tau\to\infty.
\label{eq1.8}
\end{equation}
That is, when $S(\phi)$ and $\phi$ are both real, the physical expectation values 
are recovered by taking the unphysical, Langevin time to infinity

More generally, an analytically continued version of Eq.~(\ref{eq1.7}) often holds 
as long as the supersymmetric Fokker-Planck Hamiltonian, $H_{FP}$, 
formed by taking $\partial S/\partial\phi$ 
as the superpotential, has a spectrum with positive real part and a ground state
that is nondegenerate~\cite{r9,r8}. When $S(\phi)$ is complex, $H_{FP}$ can 
still have a nondegenerate ground state and a spectrum that is real and positive.
As explained in Sec.~\ref{s2}, these criteria are the
correct ones to test for convergence of expectation values for the Hamiltonians 
studied in this paper. This is also true for several other cases
studied in Refs.~\cite{rfinal,r7d,r7e,r7a,r7b,r7c}. 
This method was successful in several cases including 
statistical mechanics problems with complex chemical potentials, field-theoretic 
calculations in Minkowski space, simulations dealing with the many Fermion
problem~\cite{r9aa}, and even non-Hermitian Hamiltonians with complex eigenvalues\cite{r3}. 

The previous studies involving the complex Langevin equation focused on cases
where the physical Hamiltonians were either Hermitian, non-Hermitian with a positive 
real part, or non-Hermitian with a real part that was negative and an imaginary 
part that was small. These cases were studied because the associated Fokker-Planck Hamiltonian, 
$H_{FP}$, had eigenvalues with a positive, real part. In contrast, 
${\cal PT}$-symmetric Hamiltonians such as 
Eq.~(\ref{eq1.1}) often have a real part that is not strictly positive and an imaginary 
part that cannot be considered small. However, in the cases we have studied, $H_{FP}$ 
still possesses a spectrum that is purely real and positive.

Reference~\cite{r9a} demonstrates that the renormalized mass squared for the anharmonic oscillator,
$M_1^2=(E_1-E_0)^2$, is proportional to the first nonzero eigenvalue of the associated 
Fokker-Planck Hamiltonian. This suggests 
that the positive real part of the spectrum of $H_{FP}$ is a consequence of  the purely 
real spectrum of these ${\cal PT}$-symmetric Hamiltonians. For the zero-dimensional
${\cal PT}$-symmetric Hamiltonians studied in this paper, the eigenvalues of the
associated $H_{FP}$ are always positive and real. We believe that a 
${\cal PT}$-symmetric Hamiltonian with real, positive eigenvalues
will always lead to a real positive part for the eigenvalues of $H_{FP}$. To prove
this would be tantamount to finding the exact conditions necessary for 
a given ${\cal PT}$-symmetric Hamiltonian to possess a real spectrum, and that is an open
problem. Moreover, in Ref.~\cite{r9a} and elsewhere, Hermitian Hamiltonians 
in Minkowski space often lead to an associated $H_{FP}$ that is non-Hermitian. In contrast,
the ${\cal PT}$-symmetric, non-Hermitian Hamiltonians 
in this paper always lead to Fokker-Planck Hamiltonians that maintain ${\cal PT}$ symmetry;
this is explained in Sec.~\ref{s2}. 

\section{Complex Langevin Equation}
\label{s2}

To gain a deeper understanding of when the Langevin equation works and of its connection to
the Hamiltonian being studied, we begin with the complex Fokker-Planck equation.
Allowing $S(\phi)$ and $\phi$ to be complex, Eq.(\ref{eq1.3}) can be divided
into its real and imaginary parts and written as two coupled equations. If one 
assumes the noise is purely real and uses
the standard methods of stochastic calculus to derive Ito's formula for two
variables~\cite{r7}, one is lead to the complex Fokker-Planck equation,
\begin{eqnarray}
&&{\partial P(\phi_R,\phi_I;\tau)\over\partial\tau}=\left({\partial\over\partial\phi_R}
Re\bigg[{\partial S\over\partial\phi}\bigg]
+{\partial\over\partial\phi_I}Im\bigg[{\partial S\over\partial\phi}\bigg]+{\partial^2\over\partial\phi_R^2}
\right)P(\phi_R,\phi_I;\tau)\nonumber\\
&&\qquad\qquad\qquad \equiv O_{FP}(\phi_R,\phi_I)P(\phi_R,\phi_I;\tau),
\label{eq1.8a}
\end{eqnarray}
where $\phi_R$ and $\phi_I$ are the real and imaginary parts of $\phi$ respectively and
$S=S(\phi_R+i\phi_I)$. Eq.(\ref{eq1.8a}) defines a purely
real probability in the complex-$\phi$ plane, but apart from a few 
simple cases~\cite{r7a,r7b}, explicit constructions of $P(\phi_R,\phi_I;\tau)$ are unknown.

Now that we are evolving the Langevin equation in the complex plane,
Eq.(\ref{eq1.6}) must be modified.  The
average over the Langevin probability must be taken as an
area integral in the complex plane given by
\begin{equation}
{\langle 0|A|0 \rangle_P\over\langle 0|0 \rangle_P}
={\int{D\phi_R~D\phi_I~A(\phi_R+i\phi_I)P(\phi_R,\phi_I;\tau)}\over\int{D\phi_R~D\phi_I~P(\phi_R,\phi_I;\tau)}}.
\label{eq1.8b}
\end{equation}
Notice that $A(\phi_R+i\phi_I)$ is an analytic function, but $P$ is not in general. Understanding how 
Eq.(\ref{eq1.7}) and Eq.(\ref{eq1.8}) are satisfied is now much more complicated
because in the limit $\tau\to\infty$ one must show how an area integral becomes  a path integral and
that a real, nonanalytic function, $P$, generates the  complex, analytic function $\exp(-S(\phi))$. This 
can be achieved in a formal manner by following the approach introduced in Refs.~\cite{rfinal,r7e}. 

For the case of $ig\phi^3/3$ in zero dimensions with $m=0$, the path integral converges
when $\exp(-S(\phi))$ is exponentially damped. Expressing the complex variable in polar coordinates,
$\phi=r\exp(i\theta)$, the Stokes' regions that are traditionally chosen for ${\cal PT}$-symmetric
theories are $-\pi<\theta<-2\pi/3$ and $-\pi/3<\theta<0$, as discussed in Sec.~\ref{s0}. 
These wedges are depicted in Fig. 1. Moreover, analytical calculations for these theories
are most easily done along the contour where there is pure exponential damping
defined by $\theta=-5\pi/6$ and $\theta=-\pi/6$. This contour is the dashed line
in Fig. 1. However, any contour whose endpoints lie in the appropriate Stokes wedges is acceptable.  
For purposes of proving convergence for the Langevin expectation
values, the easiest contour to use is $\phi=\phi_R-ib$, where $b$ is any finite constant. Along this
contour, $\exp(-S(\phi))=\exp(-ig\phi^3/3)\sim (oscillatory~term)*\exp(-g\phi_R^2b)$, and is therefore
damped as $\phi_R\to\pm\infty$. This contour is the solid line in Fig. 1.

The large $\tau$ behavior of the expectation values given by Eq.(\ref{eq1.8b})
is discovered by shifting integration variables, $\phi_I\to\phi_I-b$. This shift does
not affect the endpoints of integration and has a Jacobian of one. Any analytic function 
$A(\phi_R-ib+i\phi_I)$ can be Taylor expanded about the contour, $\phi_C=\phi_R-ib$.
This allows us to express the expectation values as
\begin{equation}
{\langle 0|A|0 \rangle_P\over\langle 0|0 \rangle_P}
={\int{D\phi_R~D\phi_I~\Big(e^{i\chi}A(\phi_C)\Big)P(\phi_R,\phi_I-b;\tau)}\over
\int{D\phi_R~D\phi_I~P(\phi_R,\phi_I-b;\tau)}},
\label{eq1.8bb}
\end{equation}
where 
\begin{equation}
\chi=\phi_I {\partial\over\partial\phi_R}.
\label{eq1.8bbb}
\end{equation} 
Integrating Eq.(\ref{eq1.8bb}) by parts infinitely many times gives:
\begin{equation}
{\langle 0|A|0 \rangle_P\over\langle 0|0 \rangle_P}
={\int{D\phi_R ~A(\phi_C) P_{eff}(\phi_C,\tau)}\over\int{D\phi_R~P_{eff}(\phi_C,\tau)}},
\label{eq1.8c}
\end{equation}
where
\begin{equation}
P_{eff}(\phi_C,\tau)=\int D\phi_I~e^{-i\chi}P(\phi_R,\phi_I-b;\tau).
\label{eq1.8d}
\end{equation}
We assume that $P$ vanishes at infinity rapidly enough so that all of the boundary terms 
from the integration by parts are zero. (In the denominator of Eq.(\ref{eq1.8bb}) $\exp(-i\chi)$
can be introduced for free because all but the zeroth order term in $\chi$ integrate
to zero.) Note that $P_{eff}$ is an analytic function of
$\phi_C=\phi_R-ib$, not a function of $\phi_R$ and $b$ separately. We see this by using
$\exp(-i\chi)P(\phi_R,\phi_I-b;\tau)=P(\phi_R-i\phi_I,\phi_I-b;\tau)$ and then shifting
the integration variable $\phi_I\to\phi_I+b$, so that $P_{eff}=\int D\phi_I~P(\phi_C-i\phi_I,\phi_I;\tau)$.
As a result, Eq.(\ref{eq1.8c}) can be equivalently written as
\begin{equation}
{\langle 0|A|0 \rangle_P\over\langle 0|0 \rangle_P}
={\int_{-\infty-ib}^{\infty-ib} {D\phi_C ~A(\phi_C)P_{eff}(\phi_C,\tau)}\over\int_{-\infty-ib}^{\infty-ib}
{D\phi_C~P_{eff}(\phi_C,\tau)}}.
\label{eq1.8f}
\end{equation}

We now derive a pseudo Fokker-Planck equation for $P_{eff}(\phi_C,\tau)$. From Eqs.~(\ref{eq1.8d}) and
(\ref{eq1.8a}),
\begin{eqnarray}
&&{\partial P_{eff}(\phi_C,\tau)\over\partial\tau}=
\int D\phi_I~e^{-i\chi}{\partial P(\phi_R,\phi_I-b;\tau)\over\partial\tau}\nonumber\\
&&\quad=\int D\phi_I~O_{FP}^{eff}e^{-i\chi}P(\phi_R,\phi_I-b;\tau),
\label{eq1.8g}
\end{eqnarray} 
where $O_{FP}^{eff}\equiv\exp(-i\chi)O_{FP}(\phi_R,\phi_I-b)\exp(i\chi)$. Using the relations
\begin{eqnarray}
&&e^{-i\chi}{\partial\over\partial\phi_R}e^{i\chi}={\partial\over\partial\phi_R}\nonumber\\
&& e^{-i\chi}{\partial\over\partial\phi_I}e^{i\chi}={\partial\over\partial\phi_I}
+i{\partial\over\partial\phi_R}\nonumber\\
&& e^{-i\chi}F(\phi_C+i\phi_I)e^{i\chi}=F(\phi_C),
\label{eq1.8h}
\end{eqnarray}
it is straightforward to show that
\begin{equation}
O_{FP}^{eff}={\partial\over\partial\phi_R}
\left({\partial\over\partial\phi_R}+{\partial S(\phi_R-ib)\over\partial\phi_R}\right)
+{\partial\over\partial\phi_I}e^{-i\chi}Im[{\partial S(\phi_R-ib+i\phi_I)\over\partial\phi}]e^{i\chi}.
\label{eq1.8i}
\end{equation} 
The last term of Eq.~(\ref{eq1.8i}) is a total derivative in $\phi_I$, and therefore,
disappears from the right side of Eq.~(\ref{eq1.8g}), again assuming $P$ vanishes rapidly
at infinity. 
Since the remaining terms of $O_{FP}^{eff}$ do not depend on $\phi_I$, they 
can be pulled out in front of the integral over $D\phi_I$.
Using the fact that $\partial/\partial\phi_R=\partial/\partial\phi_C$ on an analytic function of 
$\phi_C$, Eq.(\ref{eq1.8g})
becomes Eq.~(\ref{eq1.5}) with $\phi=\phi_C$ and $P(\phi,\tau)=P_{eff}(\phi_C,\tau)$
\begin{equation}
{\partial P_{eff}(\phi_C,\tau)\over\partial\tau}
={\partial\over\partial\phi_C}\left({\partial\over\partial\phi_C}
+{\partial S(\phi_C)\over\partial\phi_C}\right)P_{eff}(\phi_C,\tau).
\label{eq1.8j}
\end{equation}
That is, there is a pseudo-Fokker-Planck equation that defines a complex
analytic function, $P_{eff}$, that is just the analytic continuation of the Fokker-Planck equation
for real variables.

For cases where $N>3$ in Eq.(\ref{eq1.1}), a similar derivation gives the same result.
The only subtlety is in choosing the correct contour. For any $N$ a contour with finite endpoints,
$(-b,-b\cot\theta_L)$ and $(-b,b\cot\theta_R)$, as defined by Eq.(~\ref{eq0.3a}), 
can be deformed into the contour
$\phi_C=\phi_R-ib$ as shown in Fig. 2. Consequently, using the methods above, an area integral over the strip
$-b\cot\theta_L\leq\phi_R\leq b\cot\theta_R; -\infty\leq\phi_I\leq\infty$ is expressible
as a path integral over the contour $\phi_C=\phi_R-ib$ plus boundary terms involving
derivatives of $P$.  As $b$ grows larger, the area integral
approaches an integral over the entire complex plane. We expect the boundary terms to approach 
zero because the 
probability of finding a particle at $\pm\infty$ should go to zero. This is seen in 
Fig. 3 (Sec.~\ref{s3}). Following the derivation for $ig\phi^3/3$, we are again led 
to Eq.(\ref{eq1.8j})

Thus, the problem of understanding the $\tau\to\infty$ behavior of expectation values has been
reduced to one that is formally identical to that for real variables. 
We now use the methods of Parisi and Sourlas
\cite{r10} who first discovered the hidden supersymmetry in classical
stochastic equations.

If we express Eq.(\ref{eq1.8j}) in terms of $p(\phi_C,\tau)\equiv P_{eff}(\phi_C,\tau)\exp(S(\phi)/2)$, 
we obtain the Schr\"odinger equation
\begin{equation}
{-\partial p(\phi_C,\tau)\over\partial\tau}=H_{FP}\;p(\phi_C,\tau)\equiv\left(-{\partial\over\partial\phi_C}+
{1\over2}{\partial S\over\partial\phi_C}\right)\left({\partial\over\partial\phi_C}+
{1\over2}{\partial S\over\partial\phi_C}\right)p(\phi_C,\tau).
\label{eq2.1}
\end{equation} 
As claimed, $H_{FP}$ is the supersymmetric Hamiltonian formed from the superpotential 
$\partial S/\partial\phi_C$. Since $S$ is ${\cal PT}$ symmetric, $\partial S/\partial\phi_C$ 
is anti-${\cal PT}$ symmetric, and $H_{FP}$ is ${\cal PT}$-symmetric,
as claimed in the last section~\cite{r10a}. Expanding 
Eq.~(\ref{eq2.1}) yields
\begin{equation}
H_{FP}=-{\partial^2\over\partial\phi_C^2}-{1\over2}{\partial^2 S\over\partial\phi_C^2}
+{1\over4}\left({\partial S\over\partial\phi_C}\right)^2.
\label{eq2.2}
\end{equation}

If the time-independent version of Eq.~(\ref{eq2.1}),
\begin{equation}
H_{FP}\Psi_k^{FP}(\phi_C)=\lambda_k\Psi_k^{FP}(\phi_C),
\label{eq2.3}
\end{equation}
is well posed and the eigenfunctions $\Psi_k^{FP}(\phi_C)$ are complete, then
\begin{equation}
p(\phi_C,\tau)=\sum_{k=0}^{\infty}a_k\Psi_k^{FP}(\phi_C)e^{-\lambda_{k}\tau}.
\label{eq2.4}
\end{equation}
Note that $\Psi_0^{FP}(\phi_C)\equiv\exp(-S(\phi_C)/2)$ is an eigenfunction of $H_{FP}$ with 
$\lambda_0=0$.
Therefore, Eq.~(\ref{eq2.4}) becomes
\begin{equation}
p(\phi_C,\tau)=Ce^{-(1/2)S(\phi_C)}+\sum_{k=1}^{\infty}a_k\Psi_k^{FP}(\phi_C)e^{-\lambda_{k}\tau}.
\label{eq2.5}
\end{equation}
Moreover, if the spectrum of $H_{FP}$ is such that 
\begin{equation}
Re[\lambda_k]>0,~k>0, 
\label{eq2.6}
\end{equation}
it follows that
\begin{equation}
p(\phi_C,\tau)\to Ce^{-(1/2)S(\phi_C)},~\tau\to\infty.
\label{eq2.7}
\end{equation}
The $\tau$ dependence of $p(\phi_C,\tau)$ has disappeared in this limit. That 
implies $dP_{eff}/d\tau=0$ and signals that the system has reached equilibrium. 
Expressing $P_{eff}(\phi_C,\tau)$ in terms of 
$p(\phi_C,\tau)$ and taking the limit $\tau\to\infty$ gives analytically continued versions of 
Eq.~(\ref{eq1.7}), and thus Eq.~(\ref{eq1.8}), in terms of $P_{eff}$. As a result,
Langevin expectation values are shown to converge to the right side of Eq.~(\ref{eq0.4})
as $\tau\to\infty$. This result is true for our zero-dimensional ${\cal PT}$-symmetric theories as 
long as the ground state is nondegenerate. There is no evidence that ${\cal PT}$-symmetric theories 
possess a degenerate ground state, so for the purposes of this paper, we will not consider
this a possibility.

Thus, if the supersymmetric Hamiltonian $H_{FP}$ formed from the 
superpotential, $\partial S/\partial\phi_C$, has a spectrum satisfying Eq.~(\ref{eq2.6}),
the Langevin method should work as a calculational procedure. Explicitly, we have shown that
analytic continuations of Eq.~(\ref{eq1.7}) and Eq.~(\ref{eq1.8}) with $P$ replaced by $P_{eff}$, 
\begin{equation}
P_{eff}(\phi_C,\tau)\to e^{-S(\phi_C)},~\tau\to\infty
\label{eq2.8}
\end{equation}
and
\begin{eqnarray}
&&{\int{D\phi_R~D\phi_I~A(\phi_R+i\phi_I)P(\phi_R,\phi_I;\tau)}\over\int{D\phi_R~D\phi_I~P(\phi_R,\phi_I;\tau)}}
={\int_{-\infty-ib}^{\infty-ib} {D\phi_C~A(\phi_C)P_{eff}(\phi_C,\tau)}\over\int_{-\infty-ib}^{\infty-ib}
{D\phi_C~P_{eff}(\phi_C,\tau)}}\nonumber\\
&&\quad\to{\int_{-\infty-ib}^{\infty-ib} {D\phi_C~A(\phi_C)e^{-S(\phi_C)}}
\over\int_{-\infty-ib}^{\infty-ib} {D\phi_C~e^{-S(\phi_C)}}},~\tau\to\infty
\label{eq2.9}
\end{eqnarray}
follow if $H_{FP}$ has a nondegenerate ground state, wave functions that are complete, 
and a spectrum with positive real part.

\section{Numerical Methods and Results}
\label{s3}

In this section we apply the complex Langevin method to
massless versions of Eq.~(\ref{eq1.1}) in zero and one dimension and calculate the same 
time one-point and two-point
disconnected Green's functions for the cases $N=3$ and $N=4$. We begin by 
proving (under certain assumptions) that Eq.~(\ref{eq2.6}) holds in zero dimensions and explaining the 
algorithms we have used to implement simulations. We then extend these results to their 
one-dimensional analogues. 

\subsection{Zero-Dimensional Theories}
\label{s3.1}

A recent paper by Dorey {\it et al.} \cite{r0.15} shows that ${\cal PT}$-symmetric 
Hamiltonians of the form,
\begin{equation}
{\cal H}=-(\partial\phi)^2-(i\phi)^{2M}-\alpha(i\phi)^{M-1},
\label{eq3.1a}
\end{equation}
where $M$ and $\alpha$ are real and boundary conditions have been chosen as in Sec.~\ref{s0}, 
have a real positive spectrum if the conditions
$\alpha<M$ and $M\geq 1$ are both satisfied \cite{r0.15}. 
(We have set $l=0$ in the Hamiltonian given by Dorey {\it et al.}.)
This proves (for $\alpha=0$, $M=N/2$) that massless versions of Eq.~(\ref{eq1.1}) have
a positive real spectrum. 

In zero-dimensional studies of Eq.~(\ref{eq1.1}) with $m=0$, Eq.~(\ref{eq2.2}) gives
\begin{equation}
H_{FP}=-{\partial^2\over\partial\phi_C^2}-{g(N-1)\over2}(i\phi_C)^{N-2}
-{g^2\over4}(i\phi_C)^{2(N-1)},
\label{eq3.1}
\end{equation}
where the contour, $\phi_C$, is within the Stokes' wedges explained in Sec.~\ref{s0} 
and used by Dorey {\it et al.}.
Making the change of variables $\phi\to(2/g)^{1/N}\phi$, Eq.~(\ref{eq3.1}) becomes
Eq.~(\ref{eq3.1a}) with $\alpha=N-1$ and $M=N-1$.  Thus, $\alpha=M$, and as explained in
Ref.~\cite{r0.15}, this implies that the $H_{FP}$ given in Eq.~(\ref{eq3.1}) has one zero 
eigenvalue, which we have already demonstrated, and that all of the remaining 
eigenvalues are real and positive. Consequently, Eq.~(\ref{eq2.6}) holds for $N\geq 2$, and this 
implies that the complex Langevin method will work. 
 
Further, the large $\phi$ behavior of the wave functions for the eigenvalue problems defined by  
Eq.~(\ref{eq3.1}) must have the form
\begin{equation}
\Psi(\phi)\sim 
\exp\bigg[\int d\phi~\sqrt{-{g^2\over4}(i\phi)^{2(N-1)}}\bigg]\sim\exp\bigg[{g\over 2N}(i\phi)^N\bigg],
\label{eq3.1b}
\end{equation}
by $WKB$.  Apart from a factor of two, the asymptotic form of the wave functions have exactly the 
same form as $\exp(-S(\phi))=\exp(g(i\phi)^N/N)$. Thus, the Stokes' wedges that define regions 
of convergence for
the path integral are exactly the same as those that demand $\Psi(\pm\infty)=0$.
This is equivalent to noting Eq.~(\ref{eq0.3}) with $N\to 2(N-1)$ is identical
to Eq.~(\ref{eq0.3a}). That is, the wedges of convergence for the path integrals defined by $\exp(-S(\phi))$
are preserved by the boundary conditions for the wave functions of the Fokker-Planck Hamiltonian.

The most general form of the Langevin equation is
\begin{equation}
{\partial\phi\over\partial\tau}=F(\phi(\tau))+\eta(\tau)
\label{eq3.2}
\end{equation}
The simplest discretization of this is Euler's method (a first-order algorithm) and is explicitly given by
\begin{equation}
\phi(j+1)=\phi(j)+h(F(\phi(j))+\eta(j))=\phi(j)+\epsilon^{2} F(\phi(j))+\epsilon\eta'(j),
\label{eq3.3}
\end{equation}
where $j$ is an index for a Langevin-time step, $h$ is the spacing in Langevin time, 
$\epsilon={\sqrt h}$, and the $h$ dependence of $\eta(\tau)$ has been made explicit
\begin{equation}
\eta'(j)=\epsilon\eta(j),\qquad\langle\eta'(j)\eta'(k)\rangle=2\delta_{j,k}.
\label{eq3.4}
\end{equation}
This form for $\eta'(\tau)$ follows from the the normalization of the Gaussian probability 
distribution because Eq.~(\ref{eq1.4}) implies 
$\eta^2(\tau)\sim\delta(0)\sim1/h$ on a lattice. For this and the second order algorithm given 
below, there are numerical instabilities for large values of $\phi$.  The worst instabilities
arise when the potential is $-g\phi^4/4$. For $-g\phi^4/4$, it was 
necessary to restrict the absolute magnitude of $\phi$ in order to avoid these instabilities. 
A typical plot of the path followed by $\phi$ in the complex plane is shown
in Fig. 3. 

One must then take $\epsilon\to 0$.
Limiting values were obtained by fitting the data with second degree and third degree polynomials in
$\epsilon$. These fits are similar to those seen in Fig.~\ref{f4}, which is for the one-dimensional case. 
Errors are calculated by collecting the simulation data in bins of a given size and computing the 
standard deviation of the means of the bins.  The maximum error as a function of bin size
is taken to be the error for the simulation. In Table~\ref{t1} the 
numerical results obtained using Euler's method are compared with exact values given in 
Ref.~\cite{r1}. The parts of the one-point and two-point disconnected Green's 
functions that are known to vanish (e.g. $Re[G_1]$) have errors larger than their magnitude as $\epsilon\to 0$. 

Euler's method is expected to converge linearly in $\epsilon$ as $\epsilon\to 0$.
Therefore, a more accurate second-order in $\epsilon$ method is 
desirable. A second-order Runge-Kutta algorithm that led to good results in previous studies
and was first developed in Ref.~\cite{r11} is
\begin{eqnarray}
&& \tilde\phi(j)=\phi(j)+\epsilon^{2}F(\phi(j))+\epsilon\eta'(j)\nonumber\\
&& \phi(j+1)=\phi(j)+{1\over2}\epsilon^2(F(\phi(j))+F(\tilde\phi(j)))+\epsilon\eta'(j).
\label{eq3.5}
\end{eqnarray}
In our studies this method is more stable numerically than Euler's method,
and therefore, allows the inclusion of more data. 
Limiting values were obtained by fitting the data with second degree and third degree polynomials in
$\epsilon$ with the linear term set equal to zero. (These fits are similar to those seen in 
Fig.~\ref{f4} for the one-dimensional case.)
The results obtained using this algorithm are compared with exact values and the result of Euler's 
method in Table~\ref{t1}. There is good agreement.

It should be noted that for the case $N=4$, $\phi(0)$ has to be chosen in the lower half 
of the complex-$\phi$ plane or else the numerical simulations are unstable.  This is in 
accord with the $WKB$ wedges needed to properly define the boundary conditions as explained in Sec.~\ref{s0}. 
This restriction also holds in one dimension for the initial configuration at $\tau=0$.

\subsection{One-Dimensional Theories}
\label{s3.2}

In this section we apply the complex Langevin method to massless versions of Eq.~(\ref{eq1.1}) 
in one dimension with $N=3$ and $N=4$. The potentials are the same as those discussed in the last 
section, but now there is a second derivative in physical time, $-\partial^2\phi(\tau,t)/\partial t^2$, 
present in all of the equations, and $\eta(\tau)$ becomes $t$ dependent.
Thus, the eigenvalue problem for $H_{FP}$ becomes a partial 
differential equation. The spectra of eigenvalue
problems for ${\cal PT}$-symmetric partial differential equations has not been studied, 
and we do not have a proof that they are real. But, following the examples of 
zero-dimensional theories, we assume the spectrum of $H_{FP}$ has a positive real part.
The results of our simulations support this assumption.

These one-dimensional theories are more computationally expensive because they give
rise to Langevin equations that are partial differential equations. Consequently, 
the second-order algorithm in the last section is useful.
For numerical simulations of one-dimensional theories, we have to introduce a physical time 
lattice and $-\partial^2\phi(\tau,t)/\partial t^2$ becomes
\begin{equation}
-{(\phi(j,l+1)-2\phi(j,l)+\phi(j,l-1))\over a^2},
\label{eq3.8}
\end{equation}
where $l$ is an index for real time and $a$ is the spacing in physical time. The algorithms 
used in the previous section are applicable here as well, but their form has changed 
slightly. The algorithms now contain Eq.~(\ref{eq3.8}) as part of $F$ and the explicit 
lattice dependence of $\eta(\tau,t)$ is such that
\begin{equation}
\eta'(j,l)=\sqrt{ha}\eta(j,l),\qquad\langle\eta'(j,l)\eta'(k,m)\rangle=2\delta_{j,k}\delta_{l,m}.
\label{eq3.9}
\end{equation}
Further, in these algorithms there is always an $\epsilon^2$ associated with 
Eq.~(\ref{eq3.8}), and thus, the simulation is unstable unless $\epsilon<<a$. However,
there are no instabilities similar to those encountered for $-g\phi^4/4$ in the
zero dimensional case. For fixed values of $a$, we compute at various values of $\epsilon$ 
and take the limit $\epsilon\to 0$, giving the expectation
values as a function of $a$.  A typical fit for this process is shown
in Fig. 4. We then take $a\to 0$ and obtain the
expectation values in the continuum limit. A fit used to extrapolate
the value of $-iG_1$ for the potential $-g\phi^4/4$ is shown in Fig. 5.
In Table~\ref{t2} the numerical results for the  expectation values are compared with numerically 
integrated quantum-mechanical values given in Ref.~\cite{r0.14}. 
We have chosen $g$ such that Eq.~(\ref{eq1.1}) corresponds with the Hamiltonians in Ref.~\cite{r0.14}.
The agreement of our results with previous work is excellent.

\section{Conclusions and Speculations}
\label{s4}

This paper and Ref.~\cite{r1} have shown that ${\cal PT}$-symmetric theories are amenable 
to the methods of quantum field theory. The previously-used Schwinger-Dyson method is very 
accurate but is difficult to apply to higher-dimensional theories. Here we show how a 
numerical method based on the complex Langevin equation can be used to obtain precise
results. We believe that this numerical method can be applied to higher-dimensional 
theories and plan to use it for future calculations. This work represents an important step 
towards a test of the physical applicability of ${\cal PT}$-symmetric Hamiltonians.

An interesting implication of this study concerns the probability and completeness of 
${\cal PT}$-symmetric theories. 
The argument for the success of the complex Langevin method in Sec.~\ref{s2} crucially 
depends upon the eigenfunctions of $H_{FP}$ being complete (\ref{eq2.4}). Similarly,
extremely accurate results in Ref.~\cite{r0.13} crucially depend on the completeness of
${\cal PT}$-symmetric eigenfunctions. The 
possible connection between the eigenvalues of the Fokker-Planck Hamiltonians
and the Hamiltonians being simulated suggests a possible connection between the eigenfunctions as well.
Perhaps proving completeness for one of these sets of eigenfunctions would imply the completeness
of the other set.

Perhaps the most important implication for ${\cal PT}$-symmetric theories is that there is an
implicit probability distribution defined by Eq.~(\ref{eq1.8a}) and seen
in Fig. 3. Moreover, this study suggests that expectation values must be interpreted as area integrals
of observables weighted by this real probability distribution.  Previous studies
of the zeros of eigenfunctions of ${\cal PT}$-symmetric theories suggest
that the zeros interlace and become dense in a narrow band in the complex-$\phi$ plane \cite{rzeros}.
Hence, it was conjectured that completeness may have to be defined in terms of area integrals.
We suspect a strong connection between the two results and speculate that a consistent
probabilistic formulation of ${\cal PT}$-symmetric theories can only be achieved in terms
of area integrals. A probabilistic interpretation of ${\cal PT}$-symmetric theories would be a 
major advance.

A recent study of the complex Langevin equation demonstrates that problems with convergence
arise when expectation values are complex but the fixed points of the Langevin equation
lie on the real axis, and as a result, the field spends most of its time on the real axis 
and away from the correct average value~\cite{r9aa}.
It was demonstrated there that this problem could be avoided by moving the fixed points
into the complex plane. In the present study the expectation values are pure real or pure imaginary.
The fixed point of the Langevin equation is zero, which is on the real axis, but the simulations
converged without moving the fixed point into the complex plane. This seems to be because the path of the
deterministic equation ($\eta =0$) is given by
\begin{equation}
\phi(\tau)={-i \over [(N-2)(\tau+C)]^{{1\over N-2}}},
\label{eq4.1a}
\end{equation}
where $C$ is an arbitrary constant given by the initial condition. 
The field is attracted to the imaginary axis even
though the fixed point is on the real axis. Thus, it appears to be crucial that the deterministic path to
or between the fixed points is in the complex plane, but not that the fixed
points themselves lie in the complex plane. The simulations for ${\cal PT}$-symmetric theories
work in a straightforward manner; the fixed points do not need to be adjusted. We believe
that the complex Langevin equation works so nicely for ${\cal PT}$-symmetric theories because 
these theories are inherently complex. As a result, it seems that ${\cal PT}$-symmetric theories may
provide a class of toy models with new and interesting properties that are 
especially well suited for probing the intricacies of the complex Langevin equation.

The most exciting proposed application of these non-Hermitian theories is to
Higgs theories. We believe that the numerical method presented in this paper should
allow one to compute the mass of the Higgs particle in four-dimensional versions of these theories. 
Before this can be done, however, renormalization must be thoroughly understood. 
Studies of field theories in fewer than four physical dimensions represent a simpler
alternative. In particular, exact
results for the scaling exponents of an $ig\phi^3/3$ theory in two and three dimensions
can be obtained by relating them to the Lee-Yang edge singularity in one and two
dimensions respectively~\cite{rix32d,rix32d2}. Evaluation of these exponents by the
methods introduced here are in progress.

\newpage

\begin{table*}[p]
\vspace{3.0in}
\centering
\caption[]
{Numerically determined values of 
$iG_1=i\langle 0|\phi|0\rangle/\langle 0|0\rangle$ and 
$-G_2=-\langle 0|\phi^2|0\rangle/\langle 0|0\rangle$ using Euler's method and
a second-order method for zero-dimensional $ig\phi^3/3$ and $-g\phi^4/4$ theories with 
$g=1/2$. These limiting values were determined by fitting the simulated data to 
polynomials of second and third degree in $\epsilon$. For the
second-order algorithm, the term linear in $\epsilon$ is set to zero.
Exact results are listed to four significant digits in the first column.
Note that the values listed for the second-order algorithm are indeed more precise
than the results using Euler's method.}
\bigskip
\begin{tabular}{lcccccc}
\hline\hline
$N$ & $iG_1^{\rm exact}$ & $iG_1^{\rm Euler}$ & $iG_1^{\rm 2nd order}$ & $-G_2^{\rm exact}$ & $-G_2^{\rm Euler}$ & $-G_2^{\rm 2nd order}$
\\ \hline
3 & 0.9185 &   0.9198(14)   & 0.9194(7)  & 0 & - & -  \\
4 & 1.163 &   1.166(3)  &   1.164(1) & 0.9560 & 0.9623(51) & 0.9595(14)\\
\hline
\end{tabular}
\label{t1}
\end{table*}

\begin{table*}[p]
\vspace{1.0in}
\centering
\caption[]
{Numerically determined values of 
$iG_1=i\langle 0|\phi|0\rangle/\langle 0|0\rangle$ and 
$-G_2=-\langle 0|\phi^2|0\rangle/\langle 0|0\rangle$ using Euler's method and
a second-order method for one-dimensional $-g(i\phi)^N/N$ theories with $N=3$
and $N=4$. In order to compare our results with Lagrangians used in previous studies, we have set
$g=N/2$. These continuum limit values were determined by fitting the 
simulated data to third degree polynomials in $\epsilon$ for a given $a$ (for the
second-order algorithm, the term linear in $\epsilon$ is set to zero), and then 
fitting the results of those fits with second degree polynomials in $a^2$.
Exact results obtained by direct numerical integration of the quantum mechanical
problem are listed to four significant digits in the first column. 
The exact value of $-G_2$ for $N=4$ has never been directly calculated, so instead we list the 
results of the variational calculation given in Ref.~\cite{r0.14}. Note that the values listed 
for the second-order algorithm are indeed more precise than the results using Euler's method.}
\bigskip
\begin{tabular}{lcccccc}
\hline\hline
$N$ & $iG_1^{\rm exact}$ & $iG_1^{\rm Euler}$ & $iG_1^{\rm 2nd order}$ & $-G_2^{\rm var}$ & $-G_2^{\rm Euler}$ & $-G_2^{\rm 2nd order}$
\\ \hline
3 &  0.5901 &   0.5890(5)  & 0.5898(2)   & 0 & - & -  \\
4 & 0.8669 &   0.8654(6)  & 0.8670(3)   & 0.5182 & 0.5171(9) & 0.5183(4)\\
\hline
\end{tabular}
\label{t2}
\end{table*}

\begin{figure*}[p]
\epsfxsize=6.5truein
\centering\epsffile{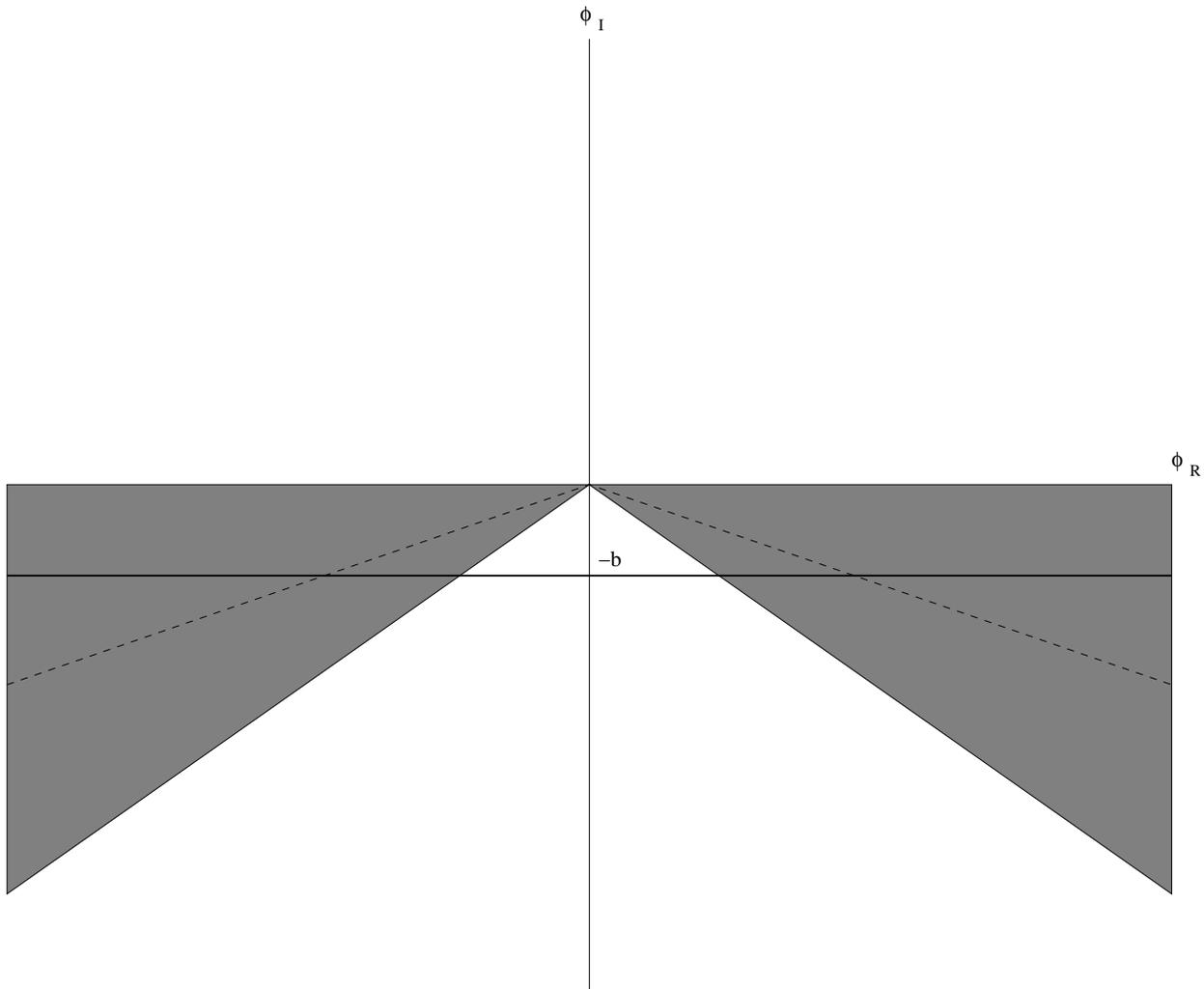}
\vspace{0.0in}
\caption
{Stokes' wedges for a massless, zero-dimensional $ig\phi^3/3$ theory.
The shaded areas are the regions within which the path integral for this theory
converges because $\exp(-S(\phi))$ is exponentially damped. The usual path of integration 
is represented by the dashed line and extends from $-\infty$ to the origin along the ray defined by 
$\theta_R=-5\pi/6$ and then from the origin to $\infty$ along the ray defined by $\theta_L=-\pi/6$. 
Along this contour $\exp(-S(\phi))$ is purely exponentially damped. In order to prove that
the Langevin expectation values have the desired behavior as
$\tau\to\infty$, it is easiest to use the smooth contour represented by the solid
line. The solid line contour is defined as $\phi=\phi_R-ib$ and extends from
$-\infty-ib\to\infty-ib$.}
\label{f1}
\end{figure*}

\begin{figure*}[p]
\epsfxsize=6.5truein
\centering\epsffile{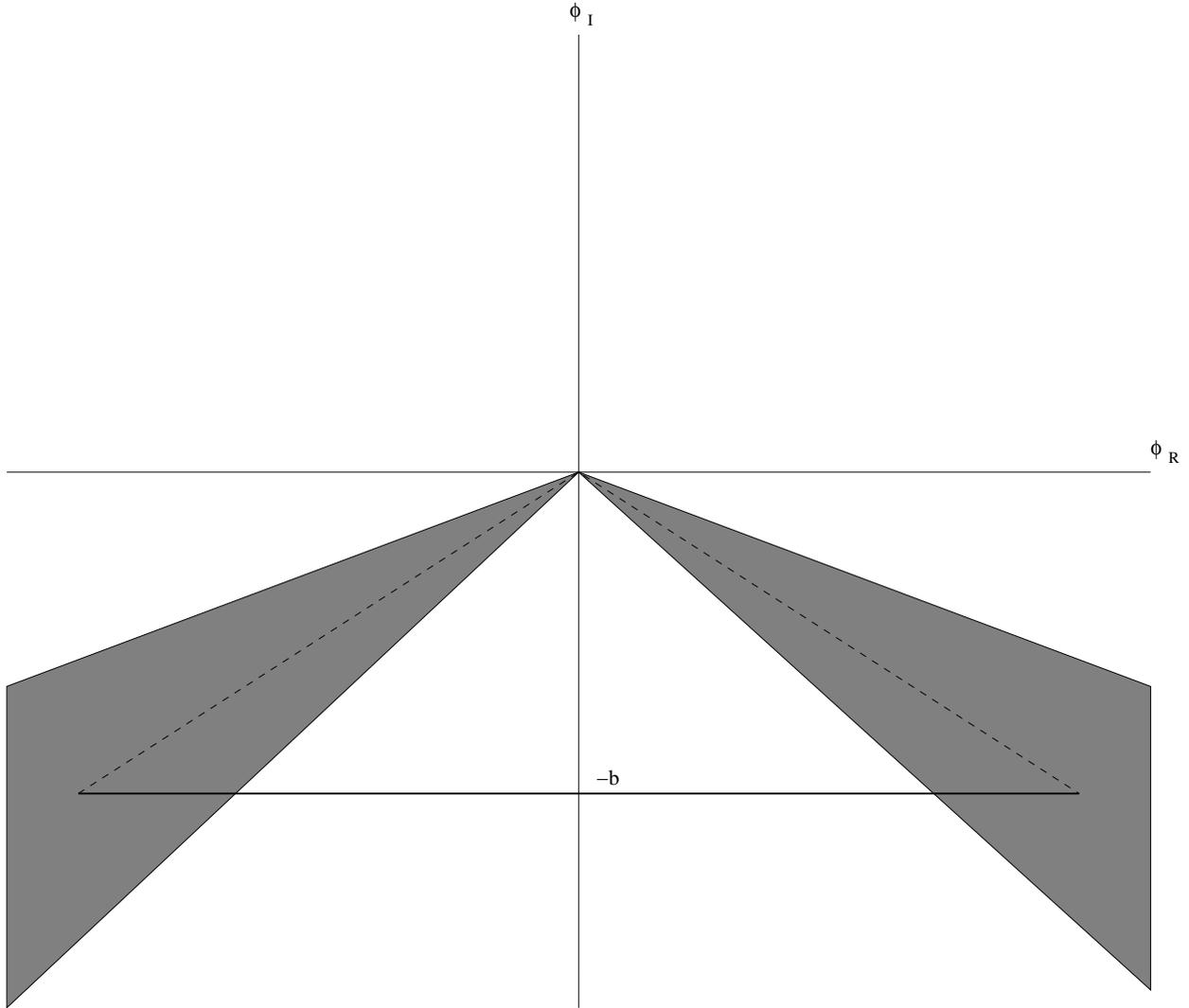}
\vspace{0.0in}
\caption
{Typical example of Stokes' wedges for a massless, zero-dimensional $-g(i\phi)^N/N$ theory.
The shaded areas are the regions within which the path integral for a given $N$
converges because $\exp(-S(\phi))$ is exponentially damped. The usual path of integration,
even for a finite contour, is represented by the dashed line and extends from $(-b,-b\cot\theta_L)$ 
to the origin and then from the origin to $(-b,b\cot\theta_R)$, where $\theta_L$ and $\theta_R$
are given in Eq.~\ref{eq0.3a}. 
In order to prove that the Langevin expectation values have the desired behavior as
$\tau\to\infty$, it is most convenient to use the smooth contour represented by the solid
line. The solid line contour is defined as $\phi=\phi_R-ib$ and extends from 
$(-b,-b\cot\theta_L)\to (-b,b\cot\theta_R)$.}
\label{f2}
\end{figure*}
\vfill\eject

\null
\vspace{6.0in}
\begin{figure*}[t]
\includegraphics{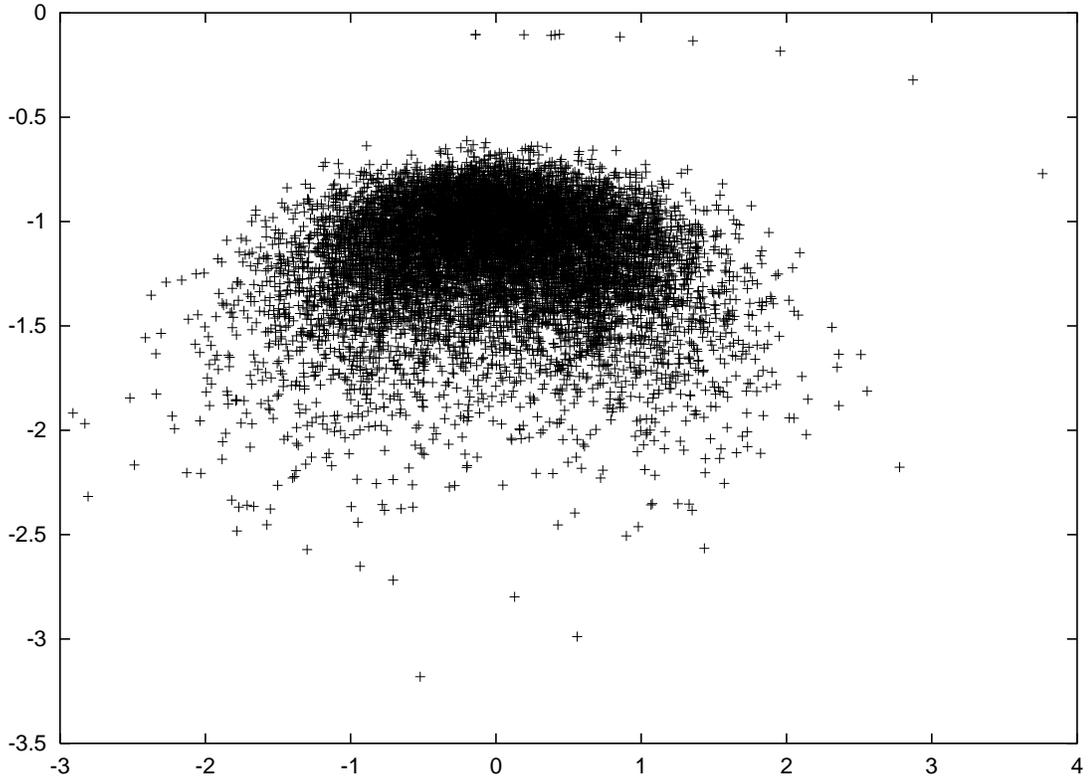}
\caption
{Plot of the field $\phi$ in the complex plane for the potential $-g\phi^4/4$ with $\epsilon=0.3$
using the second-order algorithm. Each point corresponds to the value of $\phi$ for 
a value of the fictitious time $\tau$. The absolute magnitude of $\phi$ was restricted
to be less than 19 in order avoid numerical instabilities. (For simulations of $-g\phi^4/4$ in
one dimension, this restriction was unnecessary.) The first 10,000 points are plotted.
The field started at the point $(0.5,-0.1)$ and then followed 
a path towards the imaginary axis. It then traveled from side to side forming a cloud of points
that averaged to the value $-1.1687$.}
\label{f3}
\end{figure*}

\newpage
\begin{figure*}[p]
\epsfxsize=5.5truein
\centering\epsffile{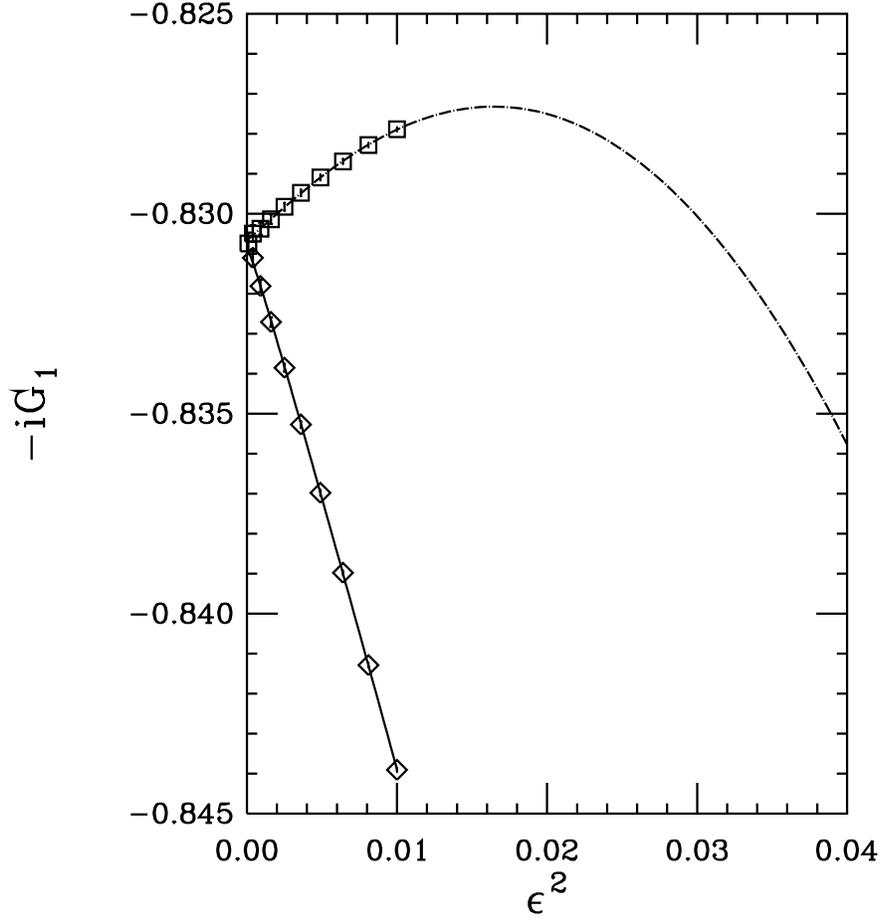}
\vskip 1.25 truein
\caption
{Plot of values for $-iG_1$ for a real time spacing of $a=0.5$ versus the 
square of the fictitious time spacing, $\epsilon^2$, for a $-g\phi^4/4$ potential in one dimension.  
The solid line is a fit that is a linear plus quadratic in $\epsilon$ for values computed using 
Euler's method (diamonds). The dashed line is a cubic fit in $\epsilon$, without a term that is linear 
in $\epsilon$, 
for values computed using the second-order algorithm (squares). Similar fits were performed for
each value of $a$ shown in Fig. 5.}
\label{f4}
\end{figure*}

\begin{figure*}[p]
\vspace{4.0in}
\epsfxsize=5.5truein
\centering\epsffile{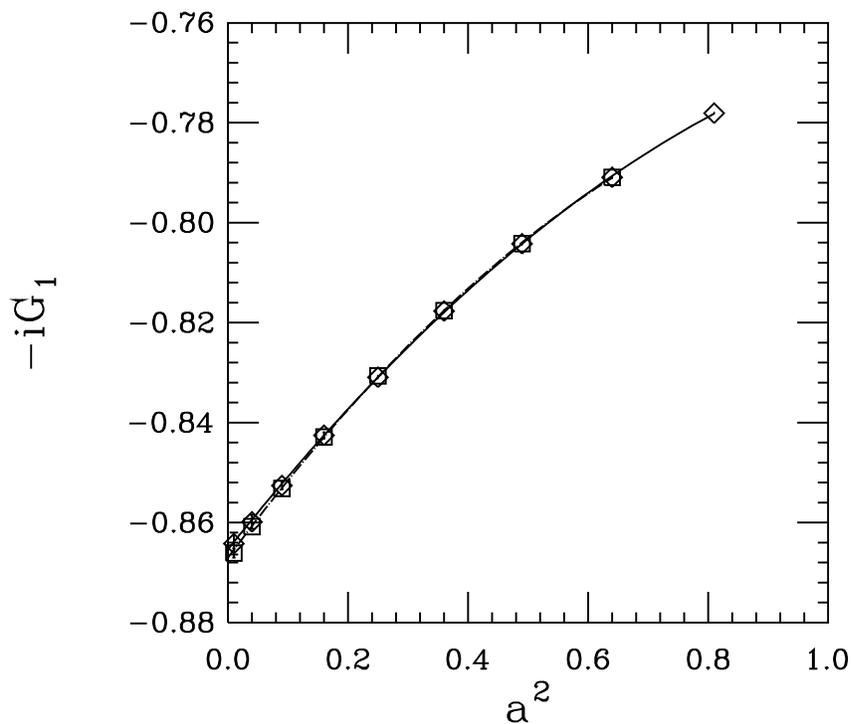}
\vskip 1.25 truein
\caption
{Extrapolated values of $-iG_1$ versus real time spacing $a^2$ for a $-g\phi^4/4$ potential in one
dimension. Lattice expectation values for these theories only depend upon even powers of $a$,
so fits are done in powers of $a^2$. Both fits shown here are linear plus quadratic in $a^2$.
Diamonds show values computed using Euler's method, and the solid line is a fit
for them. Squares show the results from the second-order algorithm; the dashed line is
the fit to them.
The continuum limit values obtained from these plots are given in Tab. 2. Similar
fits were performed for all of the data in Table 2.}
\label{f5}
\end{figure*}


\begin{references}
\bibitem{r10a} No $i$'s must be inserted to insure ${\cal PT}$ symmetry as in Ref.~\cite{r0.14}.
\end{references}
\end{document}